\documentstyle{laa}

\def\aa{{A\&A}}

\begin{document}
\input{psfig}

\thesaurus{04.01.1; 
           09.01.1; 
           }

\title{Structure Detection in Low Intensity X-Ray Images}
\author{Jean-Luc Starck and Marguerite Pierre}

\offprints{ J.L. Starck    (jstarck@cea.fr)}

\institute{CEA, DSM/DAPNIA, CE-Saclay, F-91191 Gif-sur-Yvette Cedex, France}
\date{ accepted for publication}

\maketitle

\begin{abstract}
In the context of assessing and characterizing structures in X-ray images, we compare different approaches. Most often the intensity level is very low and
necessitates a special treatment of Poisson statistics. The method based on
wavelet function histogram is shown to be the most reliable one. We also present a multi-resolution filtering method based on the wavelet coefficients
detection. Comparative results are presented by means of a simulated cluster 
of galaxies.
\end{abstract}

\keywords{Multiresolution analysis -- wavelet transform -- 
X-ray images --- image processing -- image filtering}

\section{Introduction}
The ability of detecting structures in X-ray image of celestial objects
is crucial, but the task is highly complicated due to 
 the low photon flux, typically from $0.1$ to a few photons
per pixel. Point sources
detection can be done by fitting the Point Spread Function,
but this method does not allow extended sources detection.
One way of detecting extended features in a image is to convolve it
by a Gaussian. This increases the signal to noise ratio, but at the same 
time, the resolution is degraded.  The VTP method (Scharf et al. 1997) 
allows detection of extended objects, but it is not adapted for the detection
of substructures. Furthermore, in some cases, an extended object can be 
detected as a set of point sources (Scharf et al. 1997).
The wavelet transform (WT) has been introduced (Slezak et al. 1990)
and presents considerable advantages compared to traditional methods.
The  key point is that the wavelet transform 
is able to discriminate structures as a function of scale, and thus is well suited to detect small scale structures embedded within larger scale features. Hence, WT has been used for clusters and 
subclusters analysis (Slezak et al. 1994; Grebenev et al. 1995; Rosati et al. 1995; Biviano et al. 1996), 
and has also allowed the discovery of a long, linear filamentary feature extended
 over approximatily 1 Mpc from the Coma cluster toward NGC 4911
(Vikhlinin et al. 1996). 
In the first analyses of images by the wavelet transform,
the Mexican hat was used. The method simply consists in applying
the correlation product between the image $I$ and the wavelet 
function:
\begin{eqnarray}
w_a(x,y) = I(x,y) \otimes \psi( \frac{x}{a}, \frac{y}{a})
\end{eqnarray}
Where $a$ is the scale parameter. By varying $a$, we obtain a 
set of images, each one corresponding to the wavelet coefficients
of the data at a given scale. The wavelet function corresponding to the
Mexican hat is 
\begin{eqnarray}
\psi( \frac{x}{a}, \frac{y}{a}) = (1 - \frac{x^2+y^2}{a^2}) 
  e^{-\frac{ (x^2+y^2)}{2a^2}}
\end{eqnarray}

More recently the {\em \`a trous} wavelet
transform algorithm has been used because it allows an
easy reconstruction (Slezak et al. 1994; Vikhlinin et al. 1996). By this algorithm, an image $I(x,y)$
can be decomposed into a set $(w_1,..., w_n, c_n)$, 
\begin{eqnarray}
I(x,y) = c_n(x,y) + \sum_{j=1}^{n} w_j(x,y)
\end{eqnarray}

Several statistical models have been used in order to say 
if a X-ray wavelet coefficient $w_j(x,y)$ is significant, 
i.e. not due to the noise. 
In Viklinin et al. (1996), the detection level at a given scale
is obtained by an hypothesis that the local noise follows 
a Gaussian noise. In Slezak et al. (1994),   
the Anscombe transform was used in order to transform an
image with a Poisson noise into an image with a Gaussian noise.
Other approaches have also been proposed using  
k sigma clipping on the wavelet scales (Bijaoui \& Giudicelli 1991), 
simulations (Slezak et al. 1990; Escalera \& Mazure 1992, Grebenev et al. 1995), 
a background estimation (Damiani et al. 1996; Freeman et al. 1996),
or the histogram of the wavelet function (Slezak et al. 1993; Bury 1995).

We discuss and compare in this paper the different methods for signal
detection using the {\em \`a trous} wavelet transform algorithm
 and present how X-ray images
 can be restored even in the case of very low photon flux. 
 
\section{Detection Level Estimation in the Wavelet Space}

\subsection{Model and Simulation}
Simulations can be used for deriving the probability
that a wavelet coefficient is not due to the noise (Escalera et al. 1992).
Modeling a sky image (i.e. uniform distribution and Poisson noise) allows 
determination of the wavelet coefficient distribution and derivation of a detection threshold. 
For substructure detection in a cluster, the large structure
of the cluster must be first modeled, otherwise noise photons related by 
the large scale structure will introduce false detections at lower scales.
If we have a physical model, Monte Carlo simulations can also be used 
(Escalera \& Mazure, 1992; Grebenev et al. 1995), but this approach requires
a long computation time, and the detections will always be model-dependent.
Damiani et al. (1996), and also Freeman et al. (1996) propose to 
calculate the background from the data in order to derive the fluctuations
due to the noise in the wavelet scales. It is regretable to have to do this,
because we lose one the main advantage of the use of the wavelet transform,
which is to be background-free. Indeed, wavelet coefficients have a 
null mean, and the detection is just done by comparison to a given 
threshold. Furthermore, background estimation is not an easy task, and
generally requires several steps (filtering, interpolation, etc), and 
error estimation on the background is generally difficult to calculate.  

\subsection{Sigma Clipping}
A straightforward method, initially proposed by (Bijaoui \& Giudicelli 1991),
 for deriving the detection levels 
at each scale is to apply a sigma clipping at each scale.
Therefore a standard deviation $\sigma_j$ is 
estimated at each scale $j$, and  wavelet coefficients $w_j(x,y)$ 
are  considered as significant if
\begin{eqnarray}
\mid w_j(x,y) \mid > k \sigma_j
\end{eqnarray}
where $k$ is generally taken equal to 3. This method allows us to easily
detect strong features, but is certainly not optimal for detection of weak
objects. Indeed, as the noise is not Gaussian, it is
difficult to estimate the real probability of false detection
using this $k \sigma $ detection criterion. 

\subsection{Local Gaussian noise}
Vikhlinin et al. (1995) proposed to assume a Gaussian local 
noise, and to estimate the map $I_{\sigma}(x,y)$ from the 
the local background. The standard deviation $\sigma_j(x,y)$
related to a wavelet coefficient $w_j(x,y)$ is derived from
$I_{\sigma}(x,y)$ using the property of linearity of the wavelet
transform (Starck \& Bijaoui, 1994). As previously, the hypothesis
is not true, and the consequence is the same. A solution
is to use Monte Carlo simulations to set the correspondence between
the standard deviation of a wavelet coefficient and the levels
of significance (Grebenev et al, 1995), but the simulations 
must be performed for each image because the significance levels
vary strongly with the number of photons (Grebenev et al, 1995).

\subsection{Anscombe transform}
In Slezak et al. (1994) and Biviano et al. (1996), the Anscombe transform
\begin{eqnarray}
t(I(x,y)) = 2 \sqrt{I(x,y) + \frac{2}{3}}
\end{eqnarray}
has been used and acts as if the data arose from a Gaussian
noise with white model, with $\sigma = 1$, under the assumption
that the mean value of $I$ is large. Simulations have shown 
(Murtagh et al. 1995) that a number of photons less than 30
per pixel introduces a bias. In X ray images, the number of
photons is often lower, and sometimes can even be equal to zero.
Using Anscombe transform in this case will introduce an over 
estimation of the noise level. To overcome this difficulty, 
the noise standard deviation can be reestimated, for instance
as in (Slezak et al 1994) i.e. by applying a sigma clipping at 
the first scale of the wavelet transform. However,
this approach assumes that the noise is homogeneous, 
which is not true. Indeed, if the number of photons per pixel
is lower that 30, the standard deviation of noise after 
Anscombe transformation, is varying strongly with the number 
of photons (Murtagh et al, 1995).

\subsection{Wavelet Function Histogram}
\label{noise_few_photons}
 An approach for very small numbers of 
counts, including frequent zero cases, has been described in Slezak et al.
(1993) and Bury (1994), for large scale clustering of galaxies.
We have adopted here the same approach to analyze X-ray images.

A wavelet coefficient at a given position and at a given scale $j$ is
\begin{eqnarray}
w_j(x,y) =  \sum_{k \in K} n_k \psi(\frac{x_k - x}{2^j} , \frac{y_k - y}{2^j})
\end{eqnarray}
where $K$ is the support of the wavelet function  $\psi$ (i.e. the box in which 
 $\psi$ is not equal to 0) and $n_k$ is the 
number  of events which contribute to the calculation of $w_j(x,y)$ (i.e.\ the 
number of 
photons included in the support of the dilated wavelet centered at ($x$,$y$)).
\index{wavelet transform}

If a wavelet coefficient $w_j(x,y)$ is due to the noise, it can be considered
as a realization of the sum $\sum_{k \in K} n_k$ of 
independent random variables 
with the same distribution as that of the wavelet function ($n_k$
being the number of photons or events used for the calculation of $w_j(x,y)$).
Then we compare the wavelet coefficient of the data to the values 
which can taken by the sum of $n$ independent variables.

\begin{figure*}[h]
\centerline{
\vbox{
\hbox{
\psfig{figure=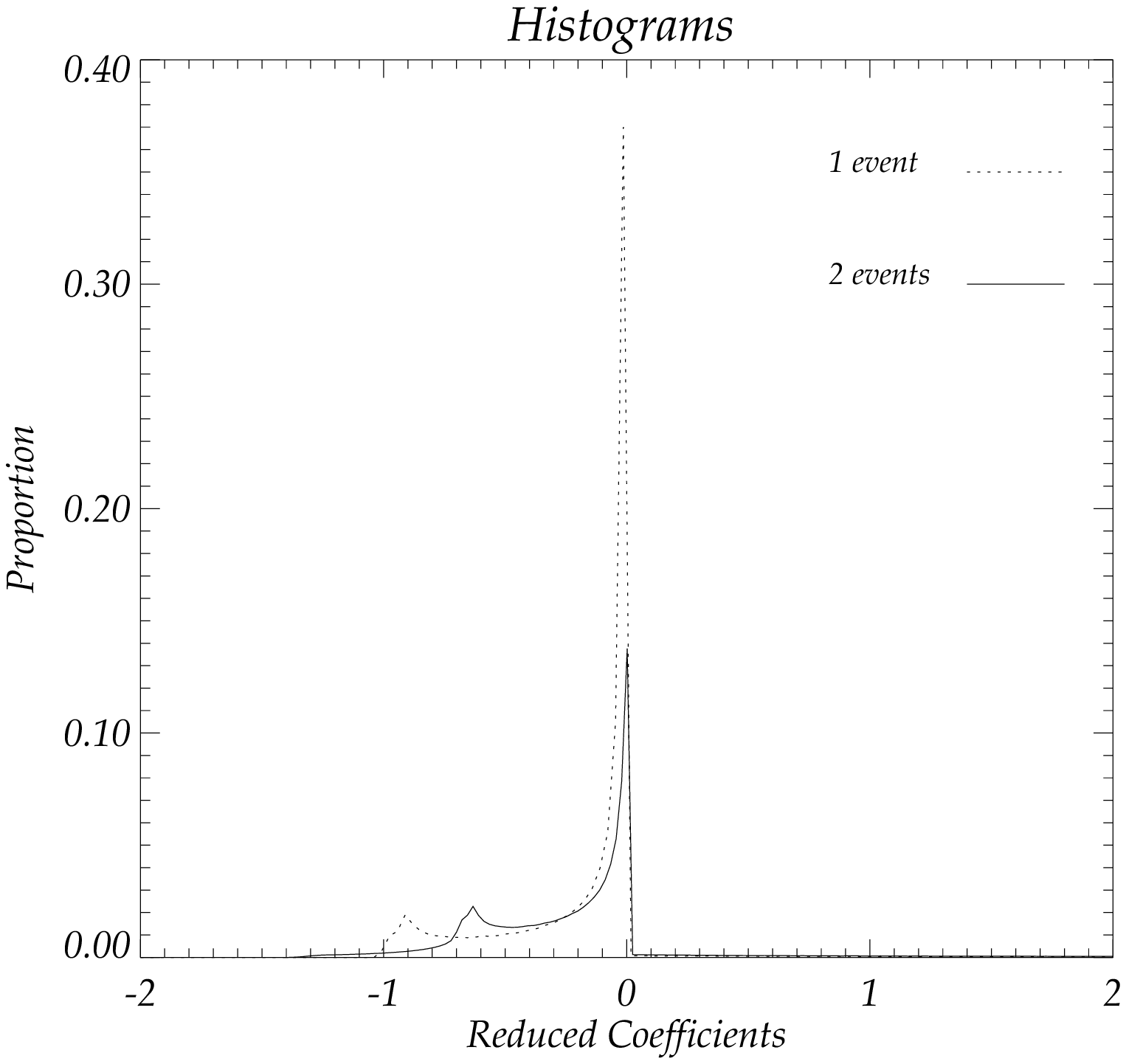,bbllx=1.5cm,bblly=1.4cm,bburx=17.5cm,bbury=16.4cm,width=7cm,height=7cm,clip=}
\psfig{figure=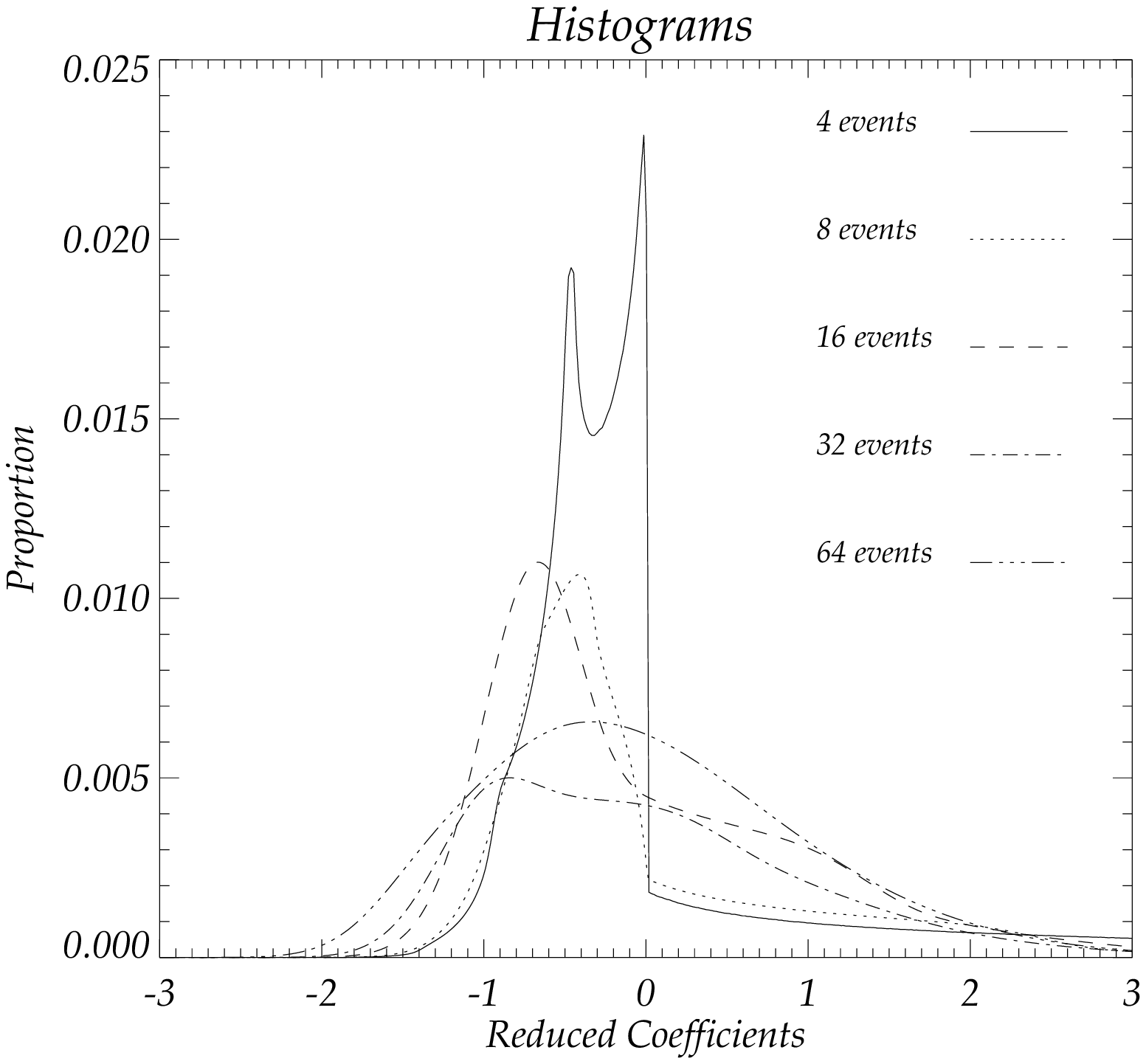,bbllx=1.5cm,bblly=1.4cm,bburx=17.5cm,bbury=16.4cm,width=7cm,height=7cm,clip=}}
\hbox{
\psfig{figure=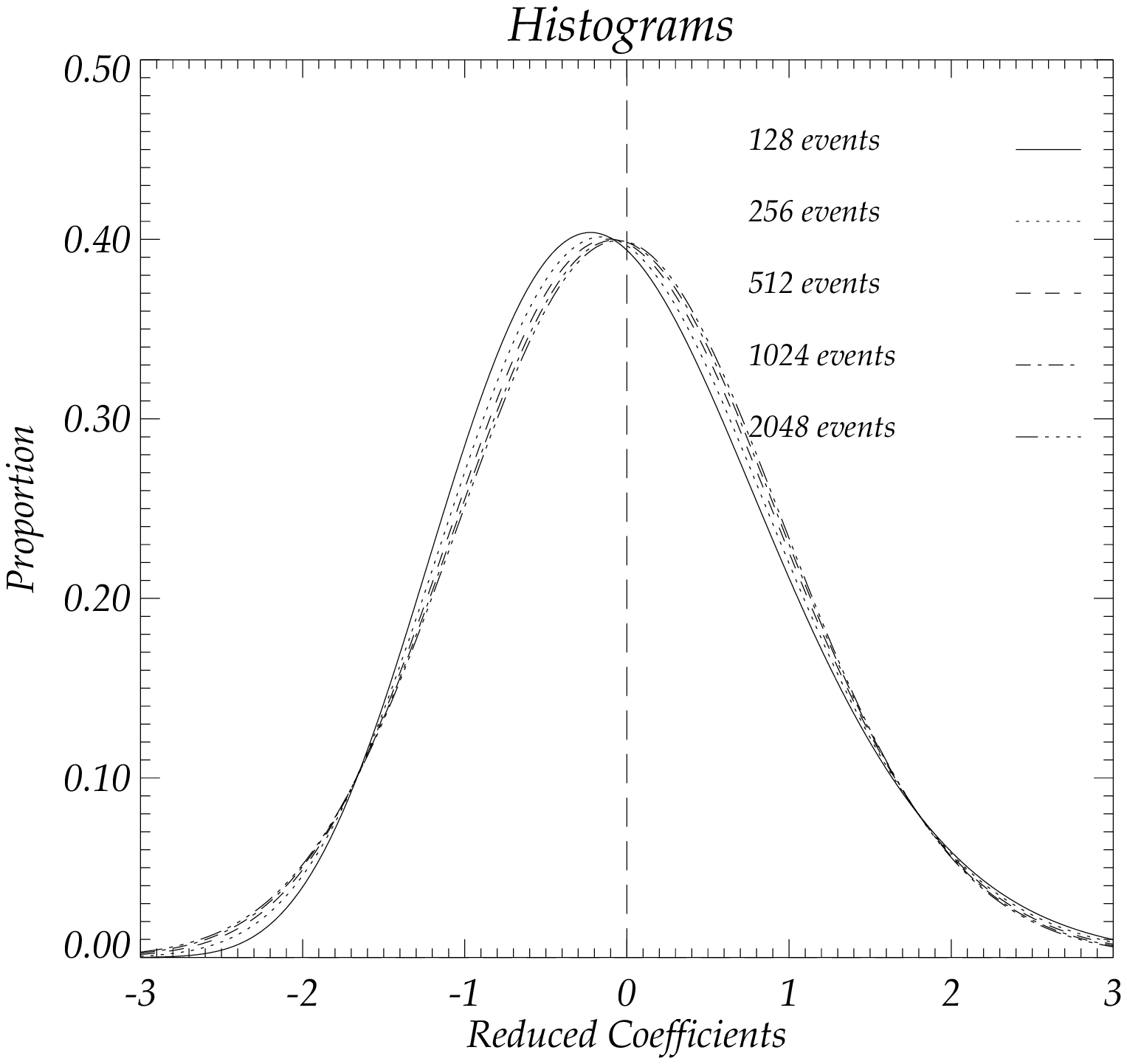,bbllx=1.5cm,bblly=1.4cm,bburx=17.5cm,bbury=16.4cm,width=7cm,height=7cm,clip=}
\psfig{figure=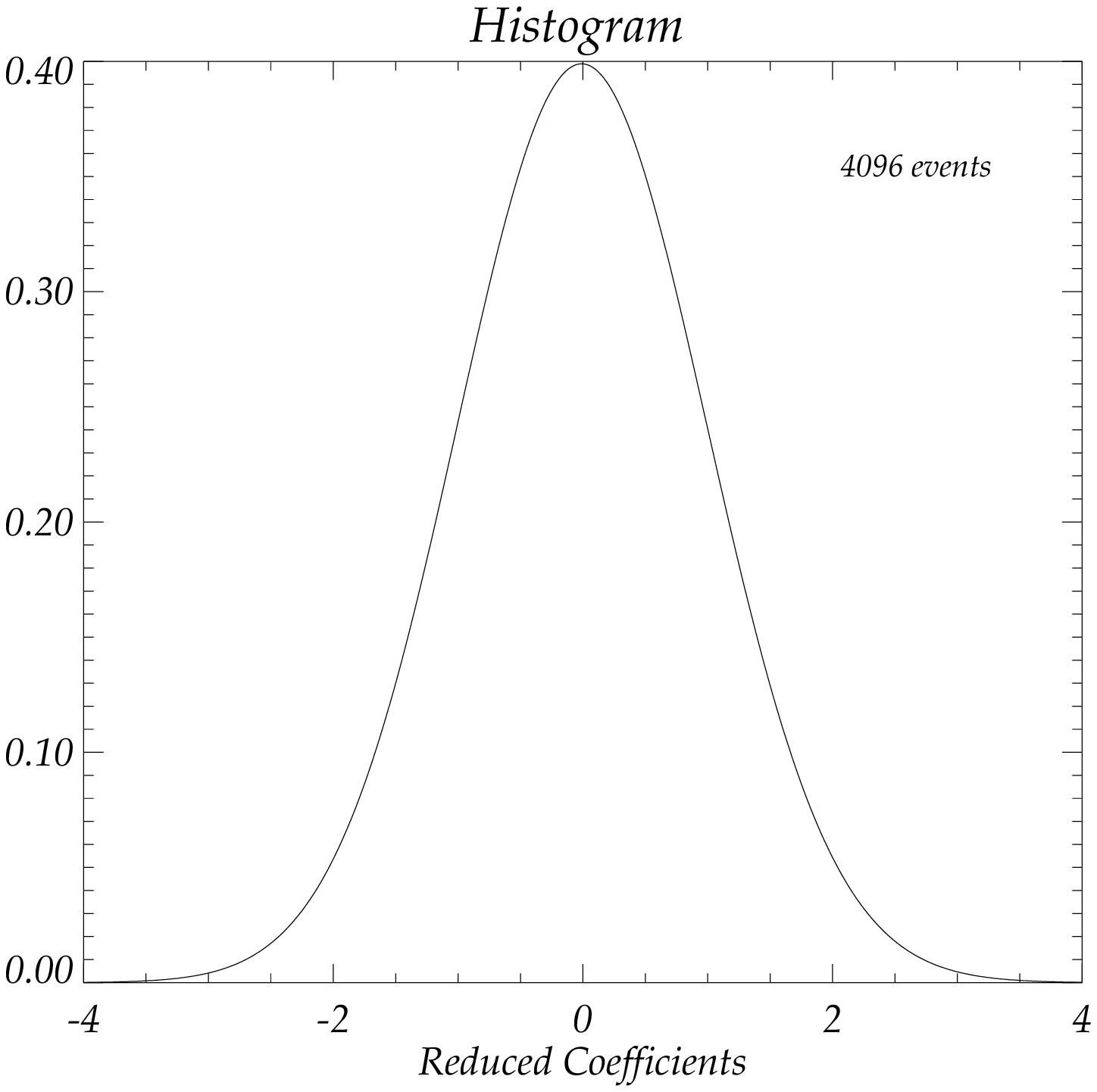,bbllx=1.5cm,bblly=1.4cm,bburx=17.5cm,bbury=16.4cm,width=7cm,height=7cm,clip=}}
}}
\caption{Autoconvolution histograms for the wavelet associated with a 
B$_3$ spline scaling function for 1 and 2 events (top left), 4 to 
64 events (top right), 128 to 2048  (bottom left), and 4096 (bottom 
right).}
\label{fig_histo}
\end{figure*}

The distribution of one event in the wavelet space is directly 
given by the histogram $H_1$ of the wavelet $\psi$. Since 
independent events are considered, the distribution of the random variable 
$W_n$ (to be associated with a wavelet coefficient) related to $n$
events is given by $n$ autoconvolutions of $H_1$
\begin{eqnarray}
H_n = H_1 \otimes  H_1 \otimes ... \otimes H_1
\end{eqnarray}
Fig.\ \ref{fig_histo} shows the shape of a set of $H_n$. For a large
number of events, $H_n$ converges to a Gaussian. 

In order to facilitate the comparisons, the variable $W_n$ of distribution
 $H_n$ 
 is reduced by
\begin{eqnarray}
c = \frac{W_n - E(W_n)}{\sigma(W_n)}
\end{eqnarray}
$E$ being the mathematical expectation, 
and the cumulative distribution  function is 
\begin{eqnarray}
F_n(c) = \int_{-\infty}^{c} H_n(u) du
\end{eqnarray}

From $F_n$, we derive $c_{min}$ and $c_{max}$ such 
that $F(c_{min}) = \epsilon$
and $F(c_{max}) = 1 - \epsilon$.

Let us define a reduced wavelet coefficient as 
\begin{eqnarray}
w^r_j(x,y)  & = &   \frac{w_j(x,y)}{\sqrt{n} \sigma_{\psi_j}} \\
             & =  & \frac{w_j(x,y)}{\sqrt{n} \sigma_{\psi}} 4^j
\label{eqn_r}
\end{eqnarray}
where $\sigma_{\psi}$ is the standard deviation of the wavelet function, 
$\sigma_{\psi_j}$ is the standard deviation of the dilated wavelet function
($\sigma_{\psi_j} = \sigma_{\psi}/4^j$), and $w_j(x,y)$ a wavelet coefficient
obtained using the {\em \`a trous} wavelet transform algorithm.

Therefore a reduced wavelet coefficient, $w^r_j(x,y)$, calculated from
$w_j(x,y)$, and resulting from $n$ photons or counts is significant if:
\begin{eqnarray}
F(w^r) > c_{max}
\label{eqn_t1}
\end{eqnarray}
or
\begin{eqnarray}
F(w^r) < c_{min}
\label{eqn_t2}
\end{eqnarray}
 
This detection method presents several advantages: it is independent of any
model, no simulation is needed, and it is theoretically rigorous.

\section{Image Filtering}
We propose here to filter an image using the multiresolution support,
which is determined from the significant wavelet coefficients
(i.e. coefficient which are not due to the noise). 

\subsection{Multiresolution Support}

A multiresolution support of an image describes in a
logical or Boolean way if an image $I$ contains information at a 
given scale $j$ and at a given position $(x,y)$.
If $M^{(I)}(j,x,y) = 1$ (or $= \ true$), then $I$ contains information at 
scale $j$ and at the position $(x,y)$.
$M$ depends on several parameters:
\begin{itemize}
\item The input image.
\item The algorithm used for the multiresolution decomposition.
\item The noise.
\item All additional constraints we want the support to satisfy.
\end{itemize}
Such a support results from the data, the treatment (noise
estimation, etc.), and from knowledge on our part of the objects contained
in the data (size of objects, linearity, etc.). In the most general case, 
a priori information is not available to us.

The multiresolution support of an image is computed in several steps:
\begin{itemize}
\item Step one is to compute the wavelet transform of the image.
\item Binarization of each scale leads to the multiresolution support
(the binarization of an image consists in assigning to each pixel a value 
only equal to $0$ or $1$). 
\item A priori knowledge can be introduced by modifying the support.
\end{itemize}
This last step depends on the knowledge we have of our images.
For instance, if we know there is no interesting object smaller or larger 
than a
given size in our image, we can suppress, in the support, anything which is
due to that kind of object. This can often be done conveniently by  the use of 
mathematical morphology. In the most general setting, we naturally have
no information to add to the multiresolution support.

The multiresolution support will be obtained by detecting 
at each scale the significant coefficients. 
The multiresolution support is defined by:
 
\begin{eqnarray}
M(j,x,y) = \left\{
  \begin{array}{ll}
  \mbox{ 1 } & \mbox{ if }   w_j(x,y) \mbox{ is significant} \\
  \mbox{ 0 } & \mbox{ if }  w_j(x,y) \mbox{ is not significant}
  \end{array}
  \right.
\end{eqnarray}

\subsection{Hard thresholding}
In the previous section, we have shown how to detect significant structures in
the wavelet scales. A simple filtering can be achieved by thresholding the
non-significant wavelet coefficients, and by reconstructing the filtered image
by the inverse wavelet transform. In the case of the {\em \`a trous} wavelet 
transform algorithm, the reconstruction is obtained by a simple addition
of the wavelet scales and the last smoothed array. The solution $S$ is
\begin{eqnarray}
S(x,y) = c_p^{(I)}(x,y) + \sum_{j=1}^{p} M(j,x,y) w_j^{(I)}(x,y) 
\end{eqnarray}
where $w_j^{(I)}$ are the wavelet coefficient of the input data, and $M$ is the
multiresolution support.

\subsection{Iterative thresholding}
As the {\em \`a trous} wavelet transform algorithm is a non orthogonal wavelet
transform algorithm, the wavelet transform of the solution $S$ does not 
produce wavelet coefficients $w_j^{(S)}(x,y)$ which are exactly 
equal to $M(j,x,y) w_j^{(I)}(x,y)$. This is evidently not a problem for wavelet coefficients where nothing was detected ( $M(j,x,y)=0$), but it means that
an error has been introduced during the reconstruction of objects from 
the significant structures. This can be corrected using an iterative method
(Starck et al., 1995). If a wavelet coefficient of the original image
is significant, then the multiresolution coefficient of the
residual image (i.e. $w^{(R^{(n)})}_j$ with $R = I - S$) must be equal to zero.
This is obtained by the following iteration:
\begin{eqnarray}
S^{k+1}(x,y) & = & S^{k}(x,y) + c_p^{(R)}(x,y) \\
            &   &  + \sum_{j=1}^{p} M(j,x,y) w_j^{(R)}(x,y)
\end{eqnarray}
Thus the regions of the image which contain significant structures at all
levels are not modified by the filtering.  The residual will contain the 
value zero over all of these regions. If an object is close to another one,
which has the same size and has a stronger flux, 
it is possible that we will not detect it because of the negative component around
the detected structure of the second object (this is due to fact that
a wavelet function has null mean). But after one or two iterations, the 
solution will contain the second object, and the residual will 
contain only the first one. This means that the wavelet coefficient (obtained
from the residual) of the first object will no longer be masked by the second. 
The multiresolution support can be updated by reducing the wavelet 
coefficient of the residual image (see~\ref{eqn_r}), and applying both 
comparison tests of eqn~\ref{eqn_t1} and eqn~\ref{eqn_t2}. Note that 
$c_{min}$ and $c_{max}$ are not recomputed, because the detection level is
unchanged.

The algorithm becomes: 
\begin{enumerate}
\item $ k \leftarrow 0 $.
\item Initialize the solution, $I^{(0)}$, to zero.
\item Determine the multiresolution support of the image.
\item Determine the residual, $R^{(k)} = I - S^{(k)}$.
\item Update the multiresolution support of the image.
\item Determine the wavelet transform $w^{(R)}$ of $R^{(k)}$. 
\item Threshold: only retain the coefficients which belong to the support.
\item Reconstruct the thresholded residual image.  This yields the image
  $\tilde{S}^{(k)}$ containing the significant residuals of the residual image.
\item Add this thresholding residual to the solution: $S^{(k)} \leftarrow S^{(k)} + 
\tilde{S}^{(k)}$.
\item If $ \mid (\sigma_{R^{(k-1)}} - \sigma_{R^{(k)}})/\sigma_{R^{(k)}} \mid \ > \ \epsilon $ then $ k \leftarrow k + 1 $ and goto 4.
\end{enumerate}
A positivity constraint can  be introduced in the algorithm by thresholding
at each iteration negative values in the solution $S$. 
The multiresolution can also be updated, following each iteration, using the wavelet coefficients of the residual image:
\begin{eqnarray}
M^{(n+1)}(j,x,y) =  \left\{
  \begin{array}{ll}
  1 & \mbox{if } w^{(R)}_j(x,y) \mbox{ is significant }\\
    & \mbox{     or } \ M^{(n)}(j,x,y) = 1  \\
0 & \mbox{if }  w^{(R)}_j(x,y) \mbox{ is not significant }   \\
  &  \mbox{      and } \ M^{(n)}(j,x,y) = 0 
  \end{array}
  \right.
\end{eqnarray}
This is of interest when an object is hidden by another one. It appends
each time a faint object is close to a stronger one. Then the faint
object is undetectable due to the negative coefficients which surrounded the 
strong one. But after one or two iterations, the strong object does 
not affect the residual, and the faint object may be appear in the scales.

\subsection{Filtering as an inverse problem}
The filtering can be seen as an inversed problem. Indeed, we want to 
reconstruct an image from the detected wavelet coefficient.
The problem of reconstruction (Bijaoui and Ru\'e 1995) consists
in searching a signal $S$ such that
its wavelet coefficients are the same as those of the detected
structure. By noting $\cal T$, the wavelet transform operator, and $P$ the
projection operator in the subspace of the detected coefficients
(i.e. set to zero all coefficients at scales and positions where
nothing where detected), the solution is found by minimization of
\begin{eqnarray}
J(S) = \parallel W - (P \circ \cal T) S  \parallel
\label{eqn_j}
\end{eqnarray}
where $W$ represents the detected wavelet coefficients of the image $I$.
A complete description of algorithms for minimization of such an equation can 
be found in Bijaoui and Ru\'e (1995). In practice, compared to the 
previous algorithm, the main modification  is the introduction of the 
adjoint wavelet transform operator, replacing the step 8 (reconstruction).

\subsection{Conclusion}
A simple thresholding  generally provides poor results. Artifacts appear 
around the structures, and the flux is not preserved. The multiresolution
support filtering requires only a few iterations, and preserves the flux.
The use of the adjoint wavelet transform operator instead of the simple
coaddition of the wavelet scale for the reconstruction (step 8 of the
algorithm) suppresses the artifacts which may appear around objects.
In fact, the algorithm is analogous to minimizing 
the equation \ref{eqn_j}. The use of the Van Cittert algorithm for 
minimization of $J$ leads to the modified multiresolution support filtering 
method. Other approaches for the minimization can also be used (conjugate
gradient, etc). The Van Cittert algorithm is not optimal for the time
computation, but it has the advantage of allowing us to add   
constraints during the iterations. The positivity is a strong constraint 
which should be used. 
Other additional  prior knowledge can be added.  For instance,
such prior knowledge could be in the form of a star position catalog, 
bad pixel positions, a given position where we expect the object to be
located, or constraints on the size of the object.  Hence the multiresolution
constraint allows us to integrate into the same data structure other
information sources (catalogs, images, etc.) and prior knowledge 
(positions, object sizes, etc.), in a way which facilitates subsequent 
image processing operations.
In the most general case, we do not have such prior 
information available, so the multiresolution support is computed from the
given input image and its noise properties.  

Partial restoration can also be considered. Indeed, we may want to restore
an image which is background free, objects which appears between two given
scales, or one object in particular. Then, the restoration must be performed
without the last smoothed array for a background free restoration, and
only from a subset of the wavelet coefficients for the restoration of
a set of objects (Bijaoui et Ru\'e 1995).

\section{Noise Models Comparison}
\label{sect_simu}

\begin{figure*}[h]
\centerline{
\hbox{
\psfig{figure=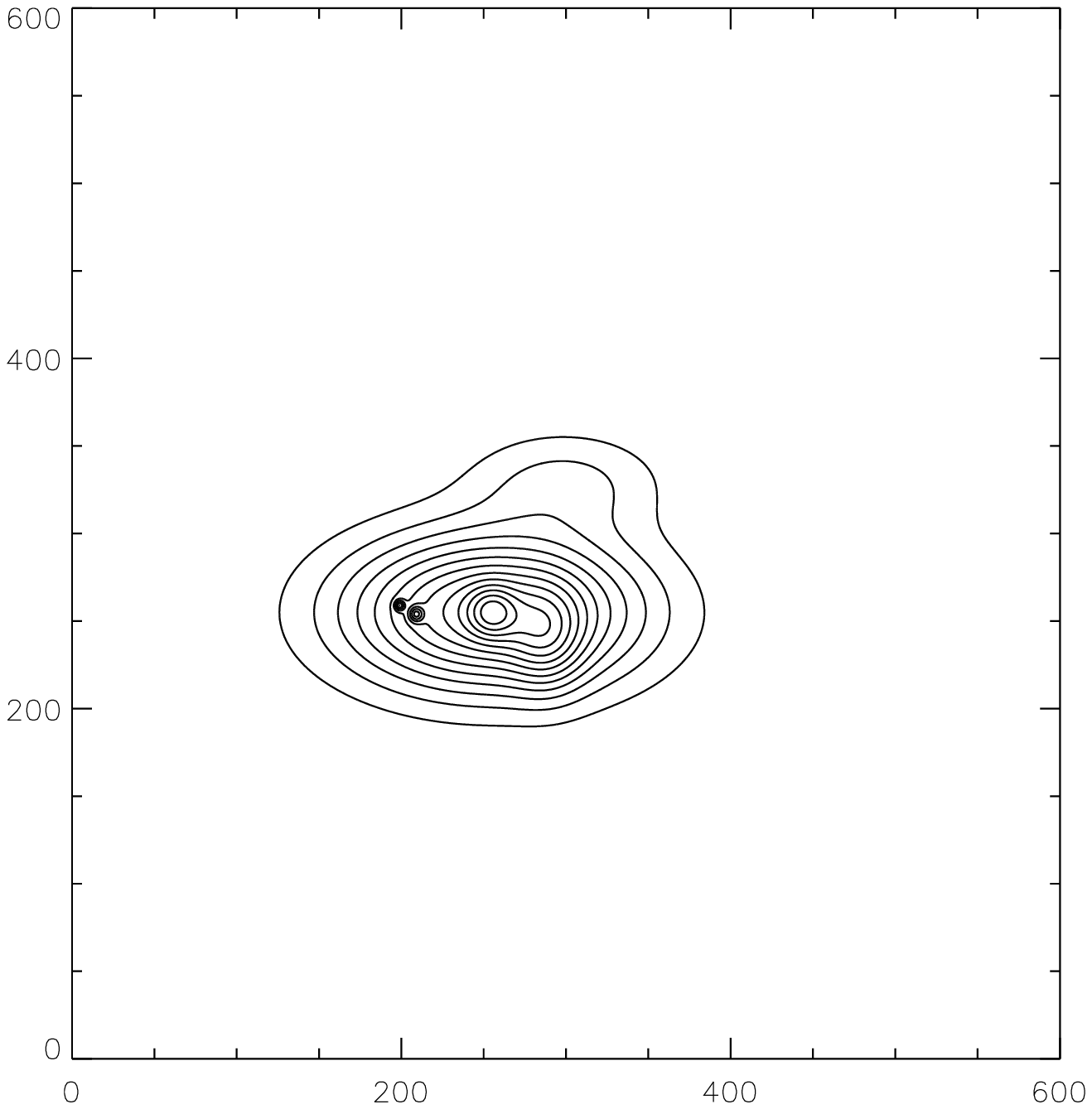,bbllx=4cm,bblly=14.cm,bburx=16.3cm,bbury=27.3cm,width=7cm,height=7cm,clip=}
\psfig{figure=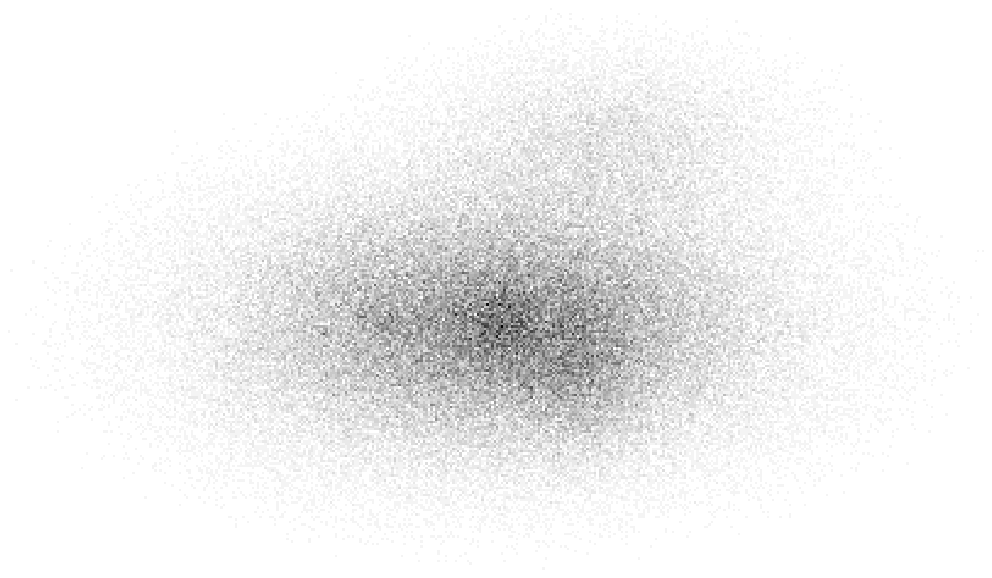,bbllx=1.8cm,bblly=12.7cm,bburx=14.8cm,bbury=25.7cm,width=7cm,height=7cm,clip=}
}}
\caption{Left, simulated image. The central luminosity is equal to 12, and the two first isophots are at $1$ and $2.62$. Right, same image with a Poisson noise.}
\label{fig_simu1}
\end{figure*}

Fig.~\ref{fig_simu1} (left) shows a simulated image of a 
galaxy cluster. Two point sources are superimposed (on the left 
of the cluster), a cooling flow is at the center, a substructure on its 
left, and a group of galaxies at the top. From this image, a 
 ``noisy'' images has been created (Fig.~\ref{fig_simu1} (right)).
The mean background level is equal to 0.1 events per pixel.
This corresponds typically to X-ray cluster observations. In the  
 noisy image, the maximum value is equal to 23 events. The background is not very relevant. The
problem in this kind of images is the small number of photons per object.
It is very difficult to extract any information from them. 

\begin{figure*}[h]
\centerline{
\vbox{
\hbox{
\psfig{figure=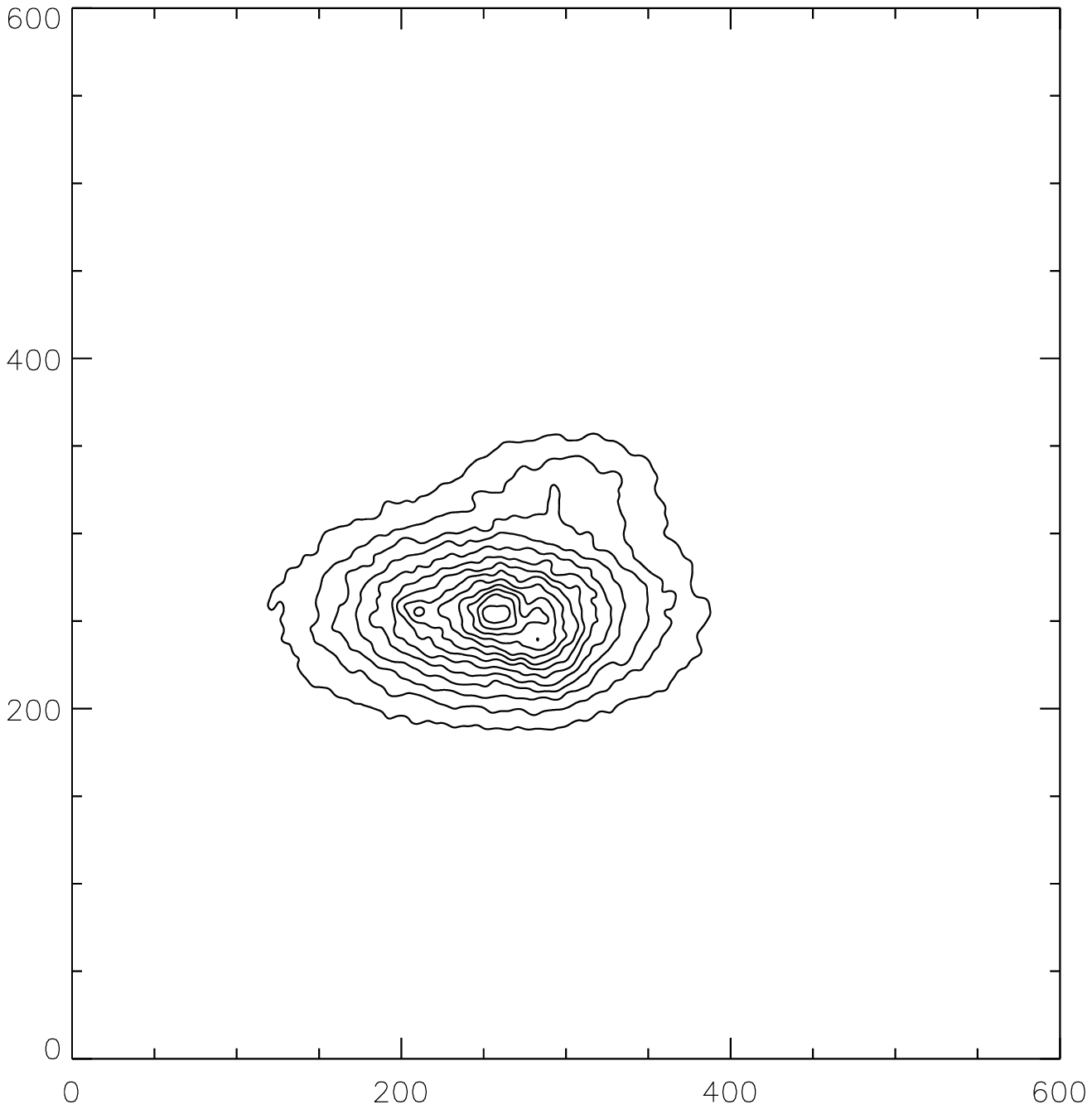,bbllx=4cm,bblly=14.cm,bburx=16.3cm,bbury=27.3cm,width=7cm,height=7cm,clip=}
\psfig{figure=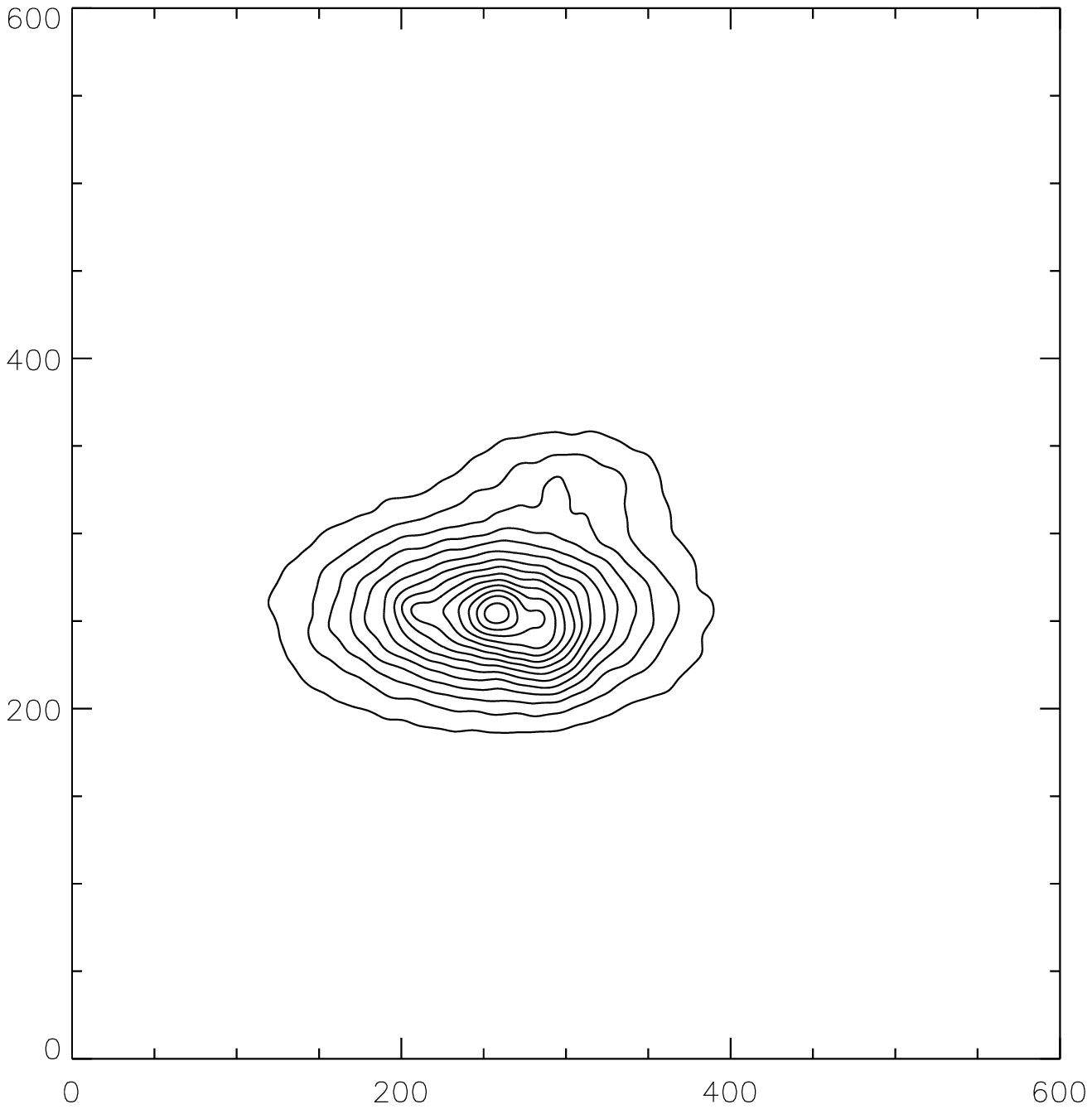,bbllx=4cm,bblly=14.cm,bburx=16.3cm,bbury=27.3cm,width=7cm,height=7cm,clip=}
}
\hbox{
\psfig{figure=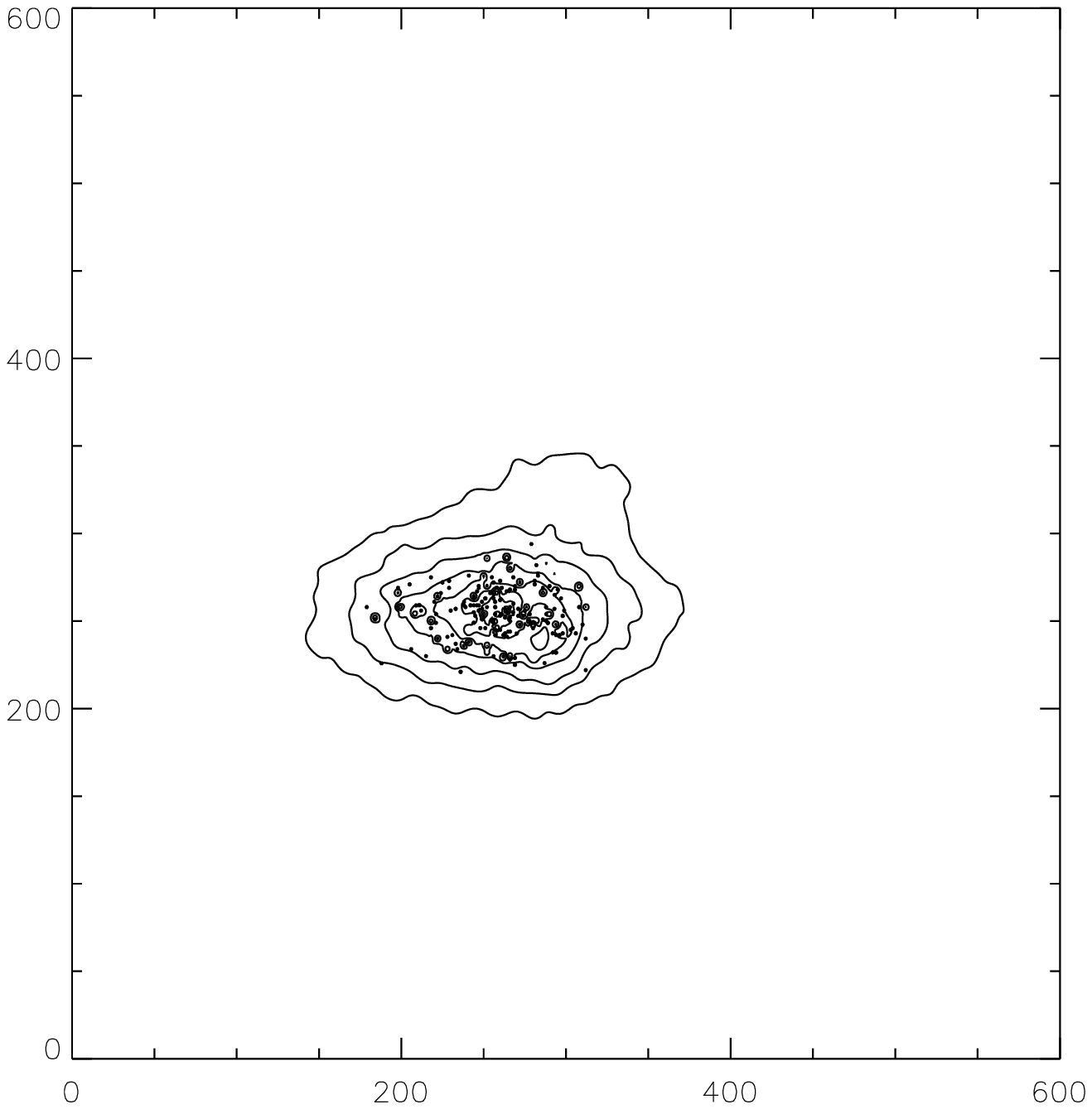,bbllx=4cm,bblly=14.cm,bburx=16.3cm,bbury=27.3cm,width=7cm,height=7cm,clip=}
\psfig{figure=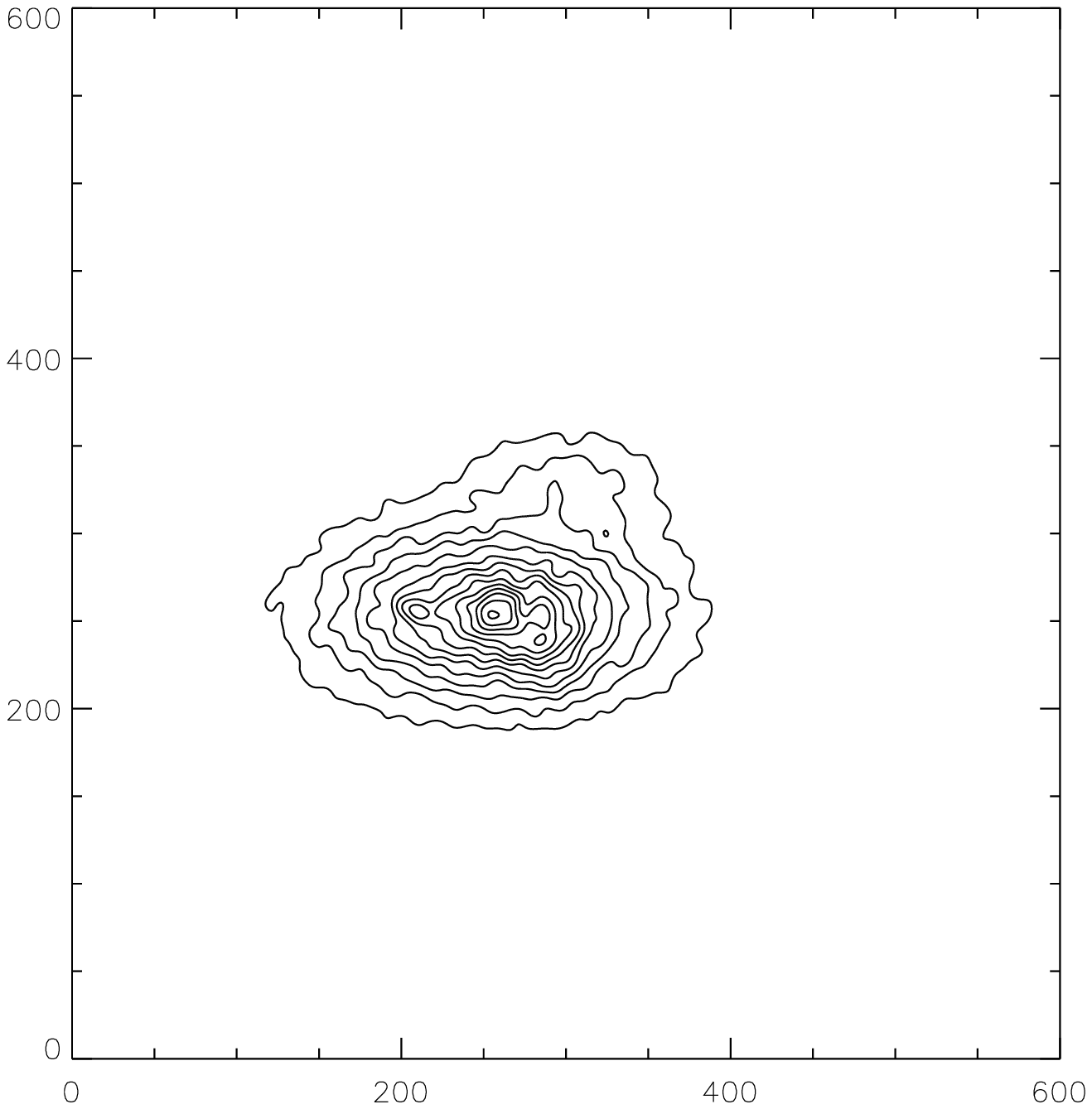,bbllx=4cm,bblly=14.cm,bburx=16.3cm,bbury=27.3cm,width=7cm,height=7cm,clip=}
}}}
\caption{Top left and right, convolution of the noisy image with a Gaussian with a standard deviation
equal to 3 and 5 respectively. Bottom left, filtered image using a sigma clipping on each wavelet scale, and a 10 sigma detection. Bottom right, filtered image using an hypothesis of local Gaussian noise, and a 10 sigma detection.  }
\label{fig_result_comp}
\end{figure*}

Fig.~\ref{fig_result_comp}, top left and top right, shows the
filtering of the image by convolving the noisy image by a Gaussian, with
a standard deviation equal to 3 and 5 respectively. 
Using the Anscombe transform, we were unable to obtain an image with a reasonable quality. It
seems that this transform should only be used in the condition defined 
in Murtagh et al (1995), i.e. with a minimum number of photons equal to
30 per pixel. In the case of very low photons count, the results are 
very poor. 

Fig.~\ref{fig_result_comp} bottom left shows the result after
a filtering using a sigma clipping on each wavelet scale, and a ten sigma detection. 
Fig.~\ref{fig_result_comp} bottom right shows the filtering using
an hypothesis of local Gaussian noise, and a ten sigma detection. For both,
even at a detection level of ten sigma, the filtered image presents 
residual noise. 

\begin{figure*}[h]
\centerline{
\vbox{
\hbox{
\psfig{figure=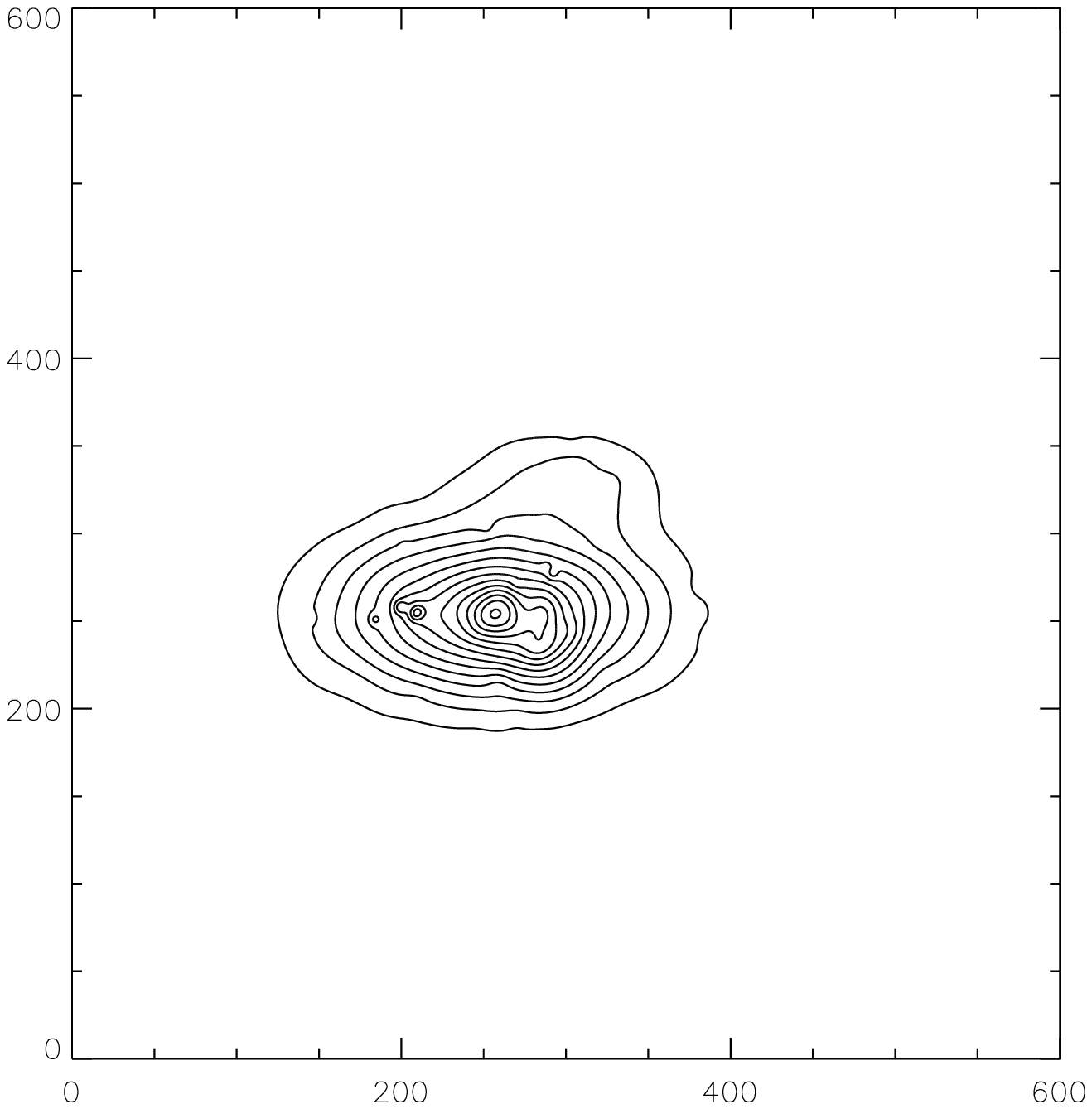,bbllx=4cm,bblly=14.cm,bburx=16.3cm,bbury=27.3cm,width=7cm,height=7cm,clip=}
\psfig{figure=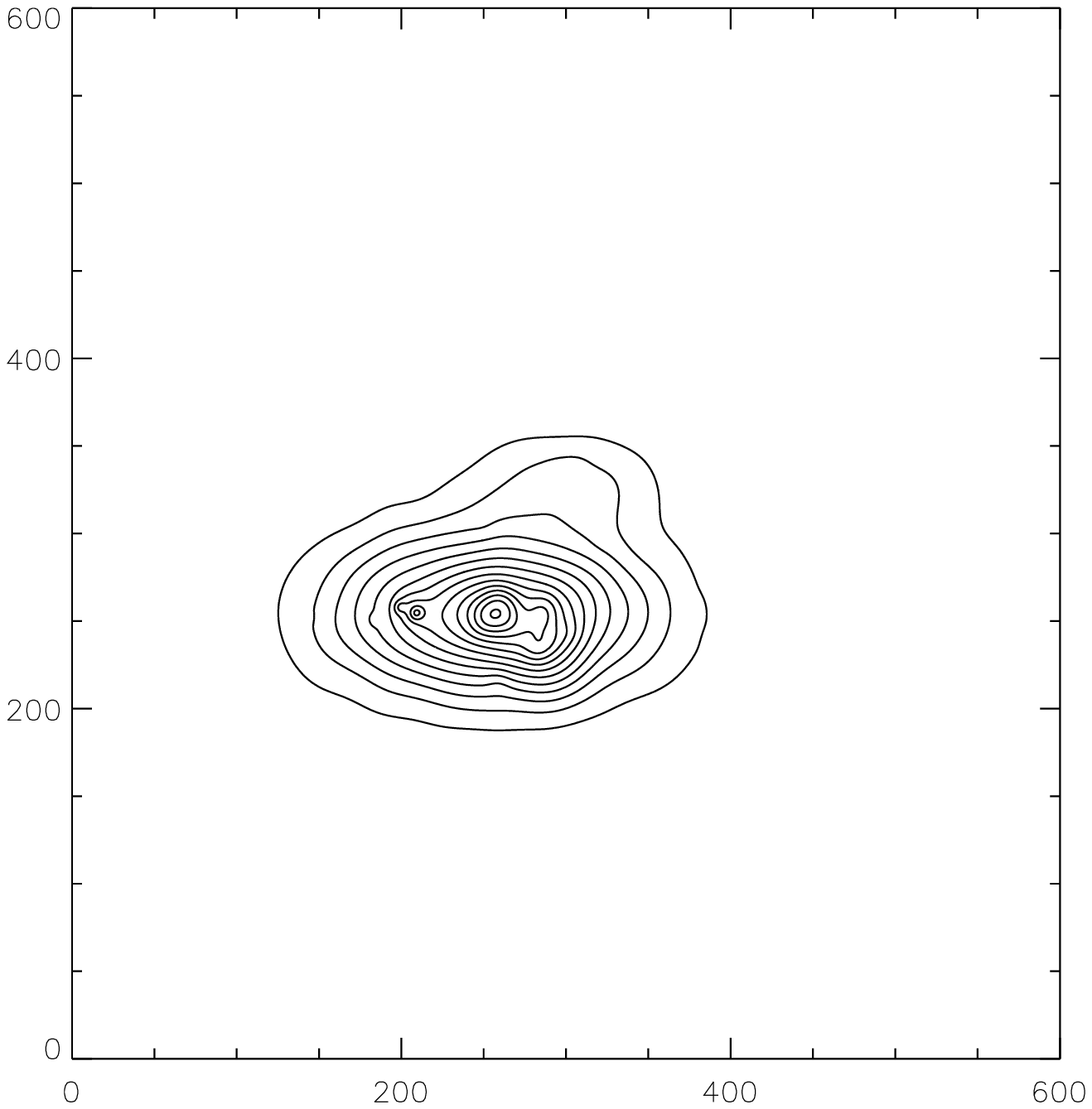,bbllx=4cm,bblly=14.cm,bburx=16.3cm,bbury=27.3cm,width=7cm,height=7cm,clip=}
}}}
\caption{Results of the filtering using the method
based on the histogram autoconvolutions. Left, image obtained with a confidence level equal to $1e-3$ (which is equivalent to a $3.09$ sigma detection for
the case of Gaussian noise), and right, image obtained with a confidence level equal to $10^{-4}$ ($3.72 \sigma$ Gaussian equivalence). }
\label{fig_result_histo}
\end{figure*}

Figure~\ref{fig_result_histo} shows the results of the filtering using the
 method based on the histogram autoconvolutions with two different 
confidence levels. Figure~\ref{fig_result_histo} left corresponds to a
confidence interval of $10^{-3}$ (which is equivalent to a $3.09$ sigma detection for
the case of Gaussian noise), and figure~\ref{fig_result_histo} right,  with a confidence level equal to $10^{-4}$ ($3.72$ Gaussian equivalence).
Even if the two point sources could not have been 
distinguished by
eye in the noisy image, they have been detected and correctly restored.

\begin{figure*}[h]
\centerline{
\vbox{
\hbox{
\psfig{figure=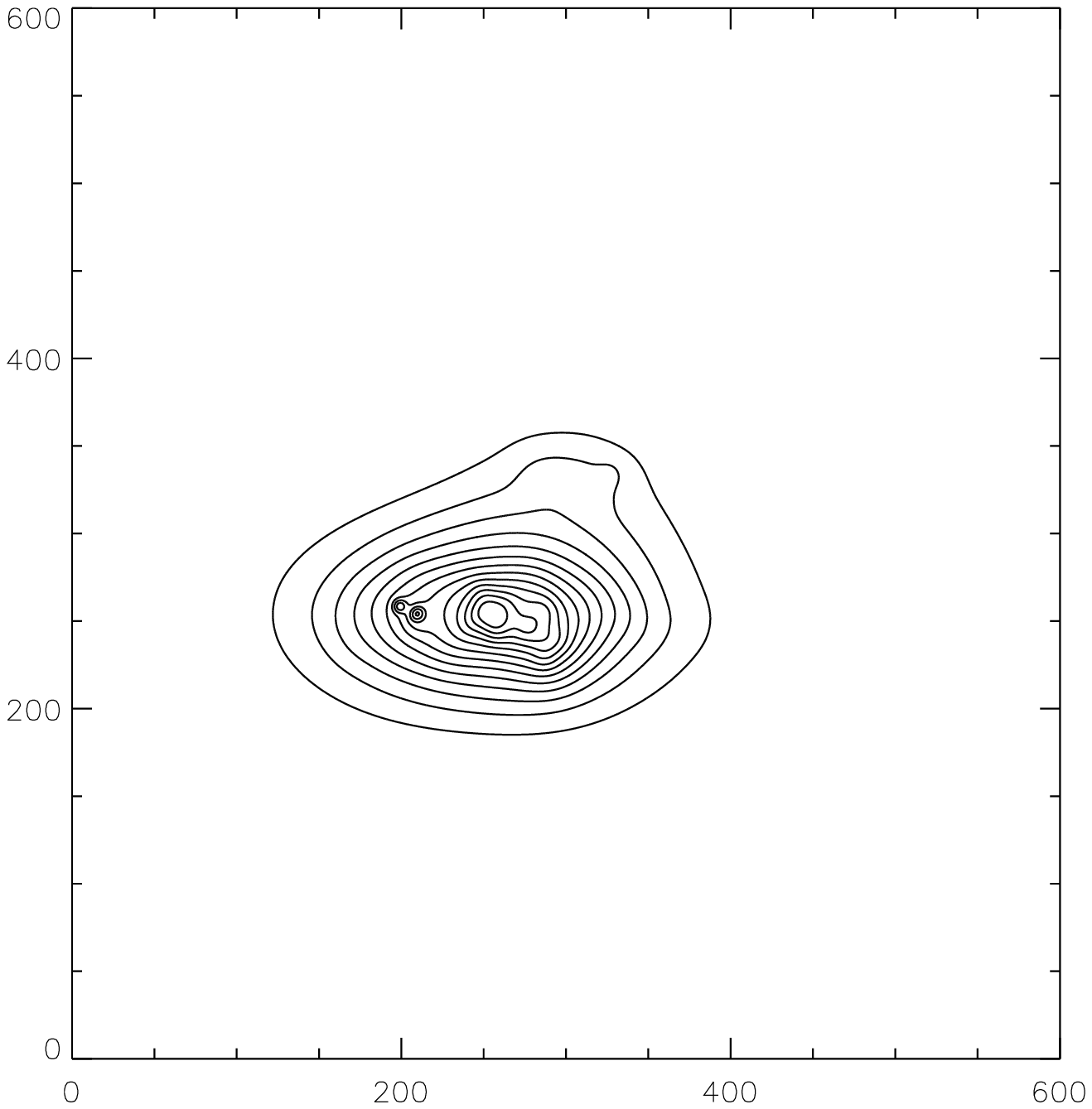,bbllx=4cm,bblly=14.cm,bburx=16.3cm,bbury=27.3cm,width=7cm,height=7cm,clip=}
\psfig{figure=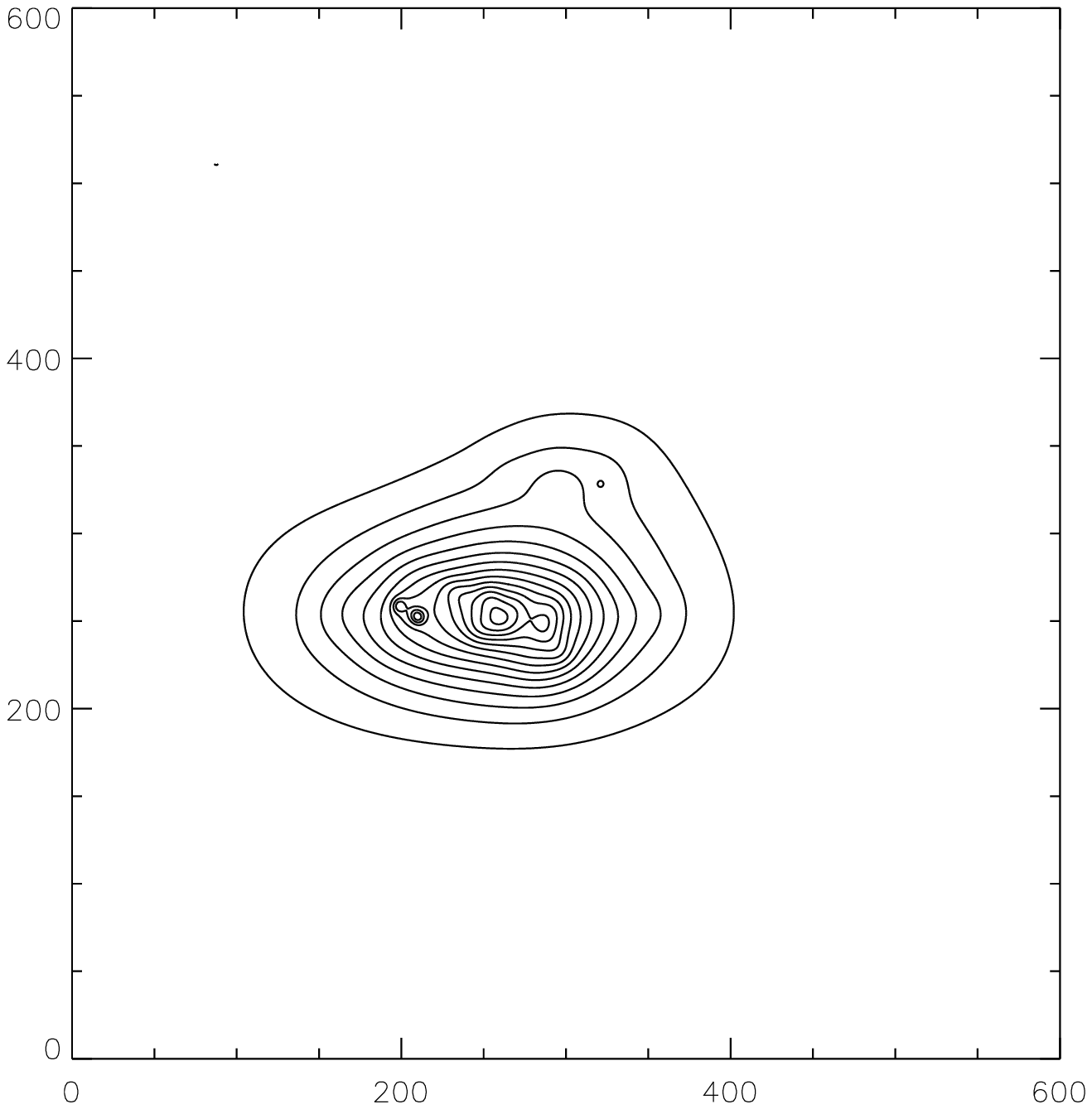,bbllx=4cm,bblly=14.cm,bburx=16.3cm,bbury=27.3cm,width=7cm,height=7cm,clip=}
}
\hbox{
\psfig{figure=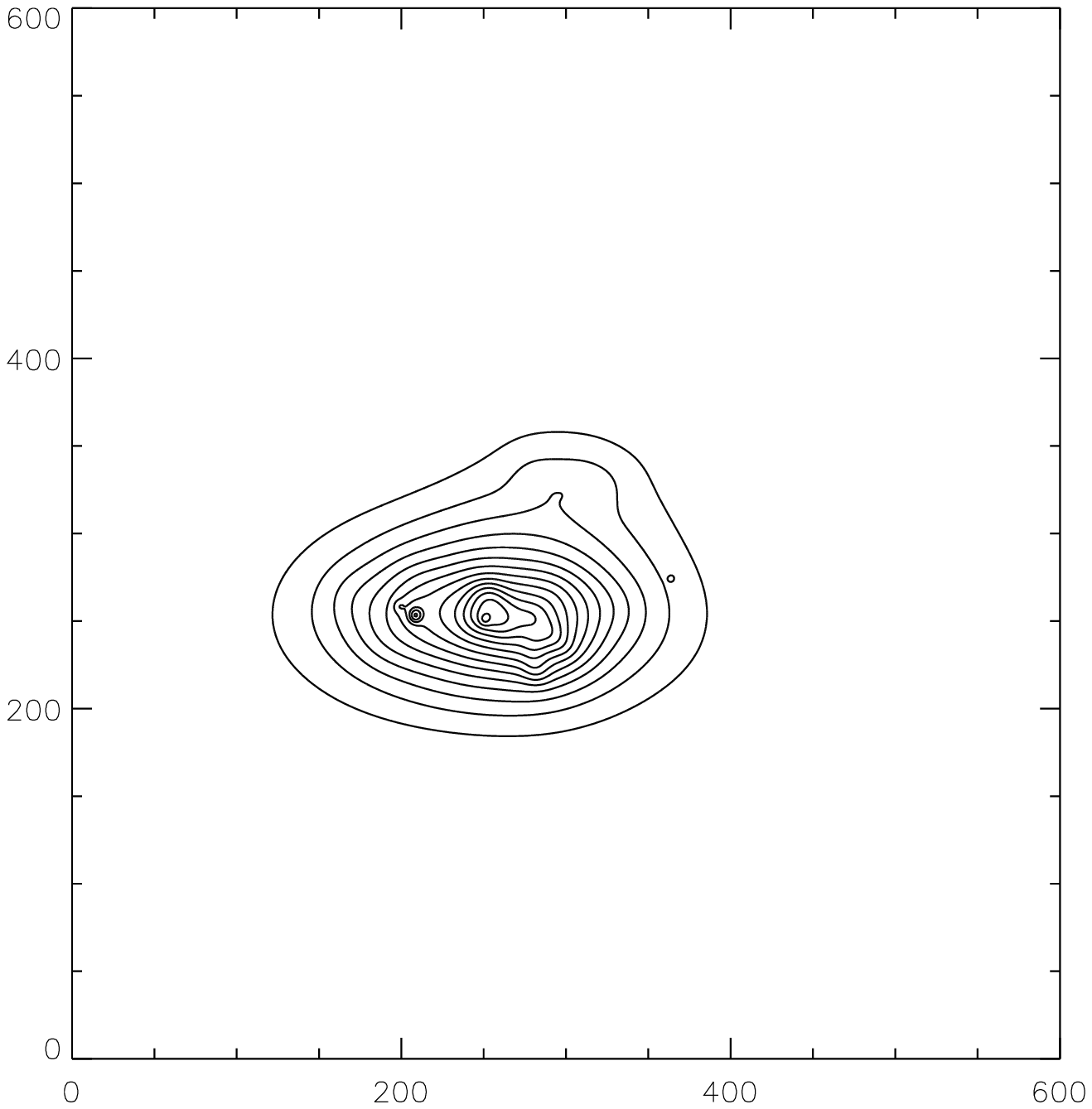,bbllx=4cm,bblly=14.cm,bburx=16.3cm,bbury=27.3cm,width=7cm,height=7cm,clip=}
\psfig{figure=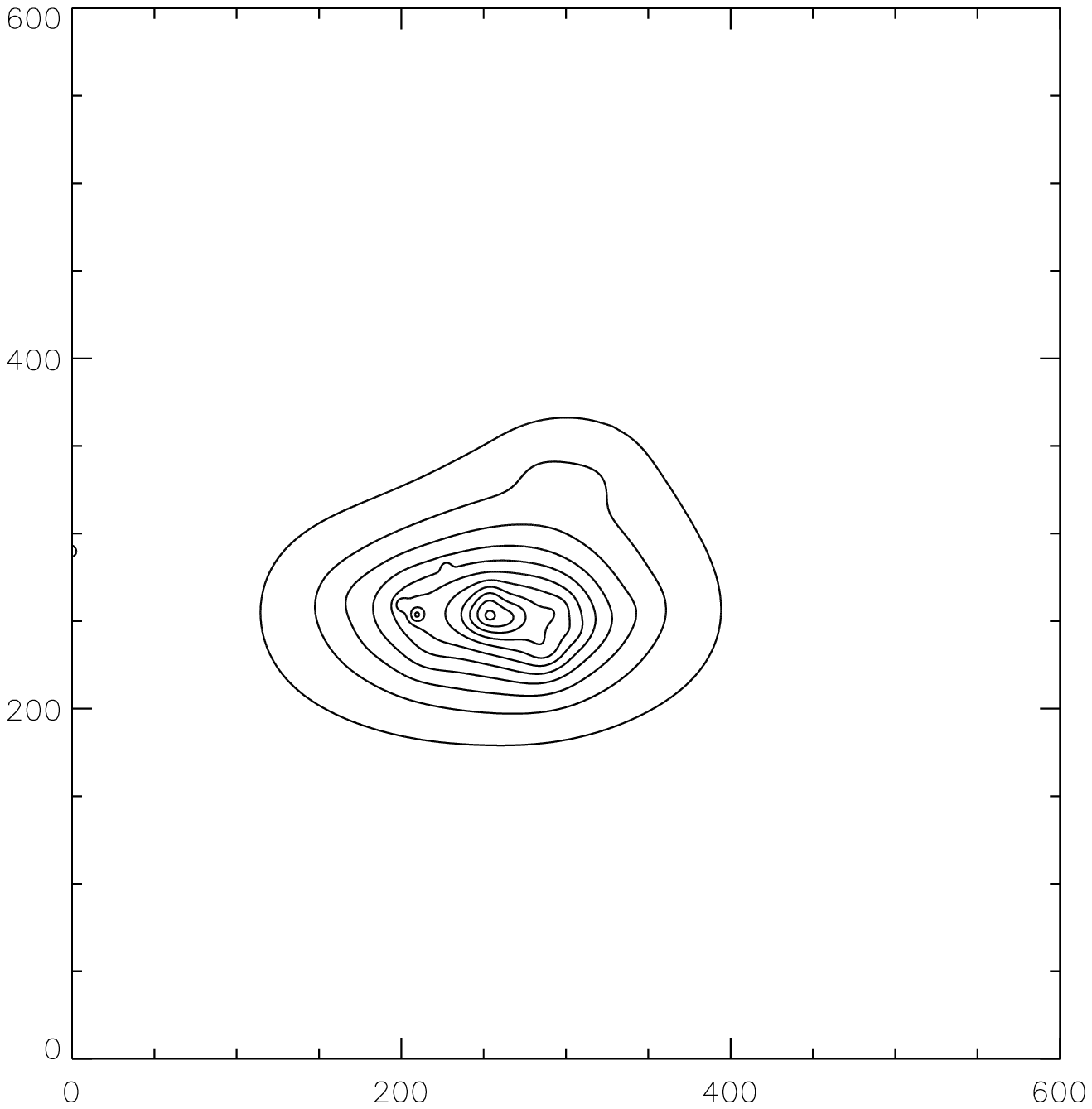,bbllx=4cm,bblly=14.cm,bburx=16.3cm,bbury=27.3cm,width=7cm,height=7cm,clip=}
}}}
\caption{Filtering of the simulated image with different background levels. 
From left to right and top to bottom, the background level was respectively
equal to $0.1$, $0.5$, $1$, $2$ counts per pixel.  }
\label{fig_result_bgr_comp}
\end{figure*}

Figure~\ref{fig_result_bgr_comp} shows the result of the filtering with
different background levels. The detections in the wavelet scale  were
 done using $\epsilon = 10^{-4}$. From left to right and top to bottom, the background level was respectively
equal to $0.1$, $0.5$, $1$, $2$ counts per pixel. If the background level
is high, there is more noise, and we see that the second source disappears
when the background level increases, which is  normal behavior. 

The best filtering is clearly obtained using the method based on wavelet 
transform and the histogram autoconvolutions. 
For other methods which use the wavelet transform, we
did not use Monte Carlo simulations and the exact level for signal detection
is difficult to find. Furthermore, the level is certainly not the same
for the whole scale. For this reason, a simple Gaussian filtering seems to be
better. 

\section{Detected Structure Analysis}
Once the significant wavelet coefficients have been detected, they can
be grouped into structures (a structure is defined as a set of connected
wavelet coefficients at a given scale), and each structure can be analyzed
independently. Interesting information which can be easily extracted
from an individual structure includes the first and second order moments, 
the angle, the perimeter, the surface, and 
 the deviation of shape from sphericity (i.e. $4 \pi \frac{Surface}{Perimeter^2}$).
From a given scale, it is also interesting to count the number structures,
and the mean deviation of shape from sphericity.

\begin{figure*}[h]
\centerline{
\vbox{
\hbox{
\psfig{figure=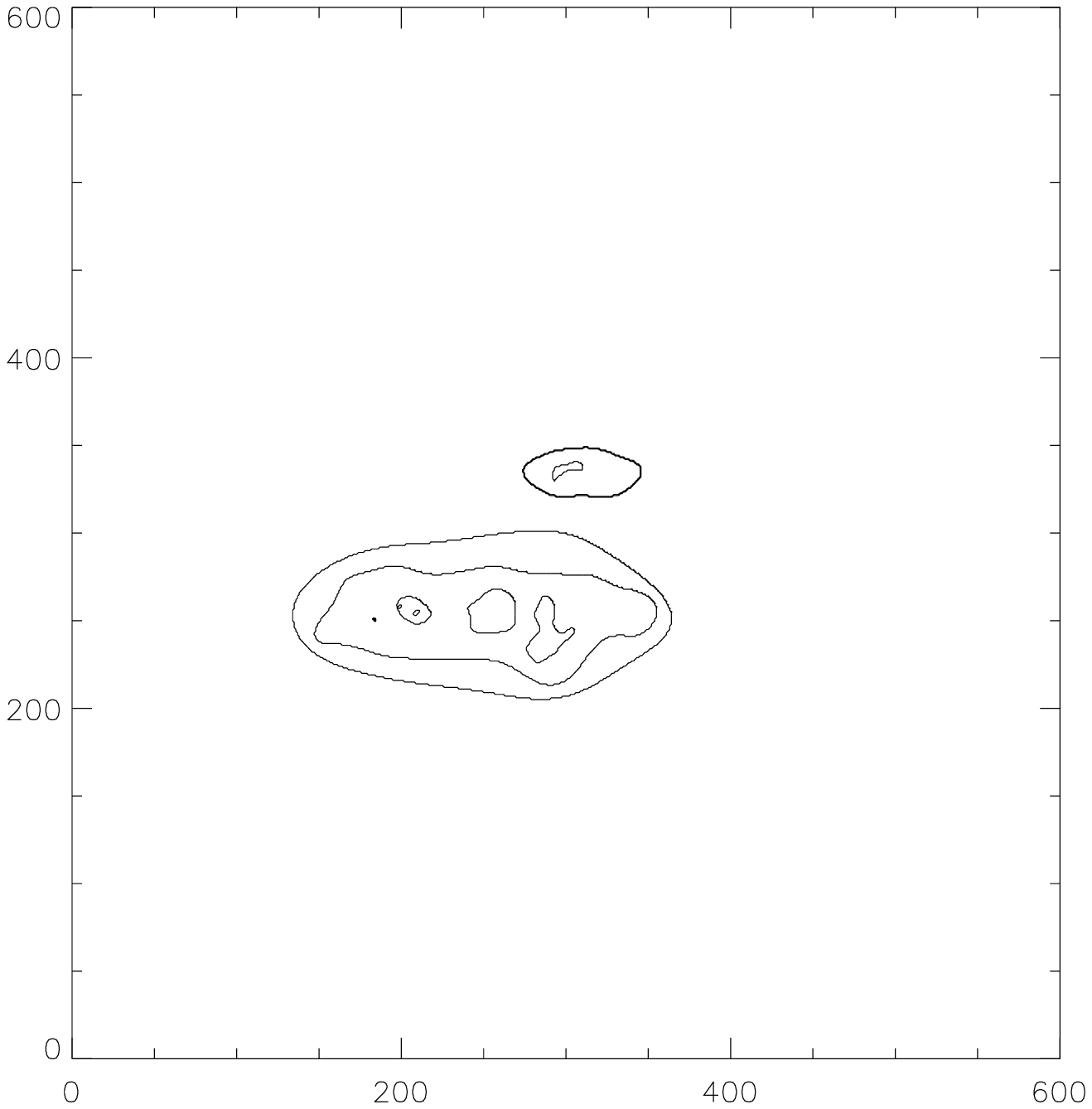,bbllx=4cm,bblly=14.cm,bburx=16.3cm,bbury=27.3cm,width=7cm,height=7cm,clip=}
\psfig{figure=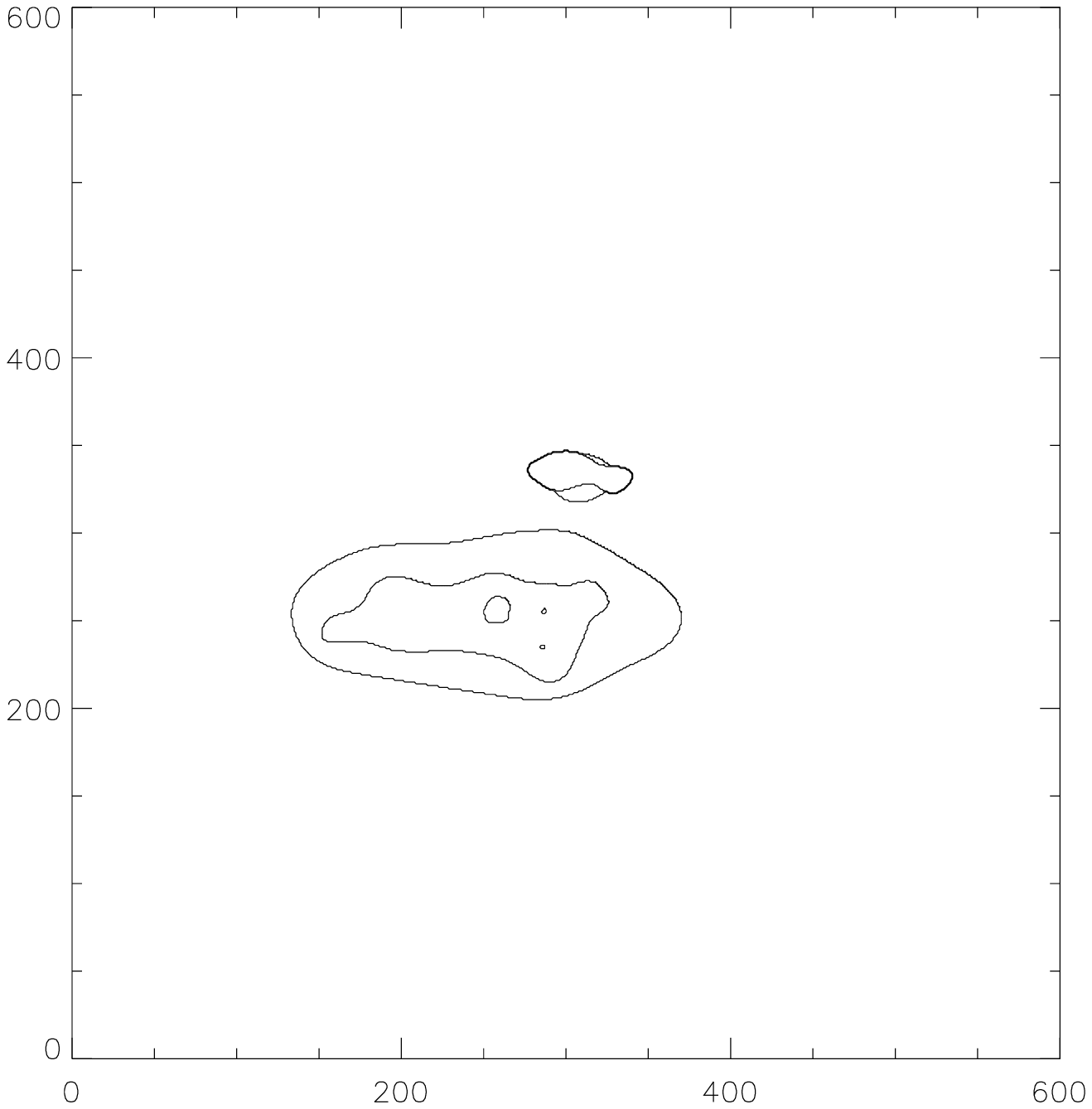,bbllx=4cm,bblly=14.cm,bburx=16.3cm,bbury=27.3cm,width=7cm,height=7cm,clip=}
}}}
\caption{Left, multiresolution support of the simulated image (see figure~	2). Right, multiresolution support of the same simulated field, but all objects contains less flux. The maximum of the noisy image is equal to 7.}
\label{fig_cont_support}
\end{figure*}

In order to visualize the structures, we can create an image by plotting
 a contour for each detected structure. This provides a compact way to visualize
the multiresolution support. 
Figure \ref{fig_cont_support} (left) shows the contours of the multiresolution 
support of the simulated image of Section~\ref{sect_simu}. Figure \ref{fig_cont_support} (right) shows the contours of the same simulated 
field, but the objects of the simulated noisy image contain less flux (the
maximum of the image is equal to seven counts), while the background is the
at the same level ($0.1$ count per pixel). We can easily see that in this 
case, the two point sources have disappeared. Both detection were
done with $\epsilon = 10^{-4}$.

\begin{figure*}[h]
\centerline{
\vbox{
\psfig{figure=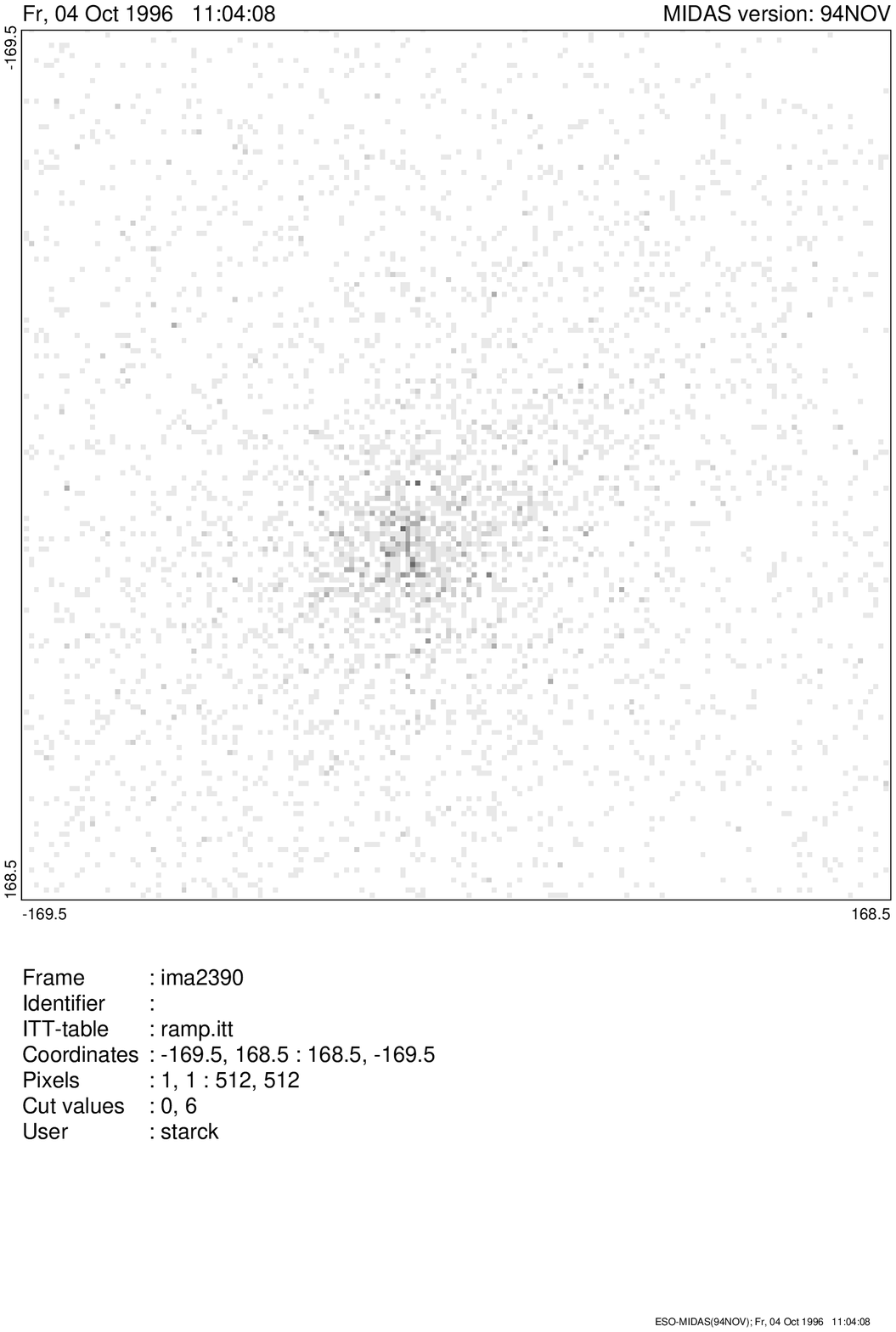,bbllx=3.1cm,bblly=10.5cm,bburx=18.7cm,bbury=26.2cm,width=10cm,height=10cm,clip=}
}}
\caption{ROSAT image of the cluster A2390.}
\label{fig_ima_a2390_bw}
\end{figure*}

\section{A2390 cluster filtering}
\begin{figure*}[h]
\centerline{
\vbox{
\psfig{figure=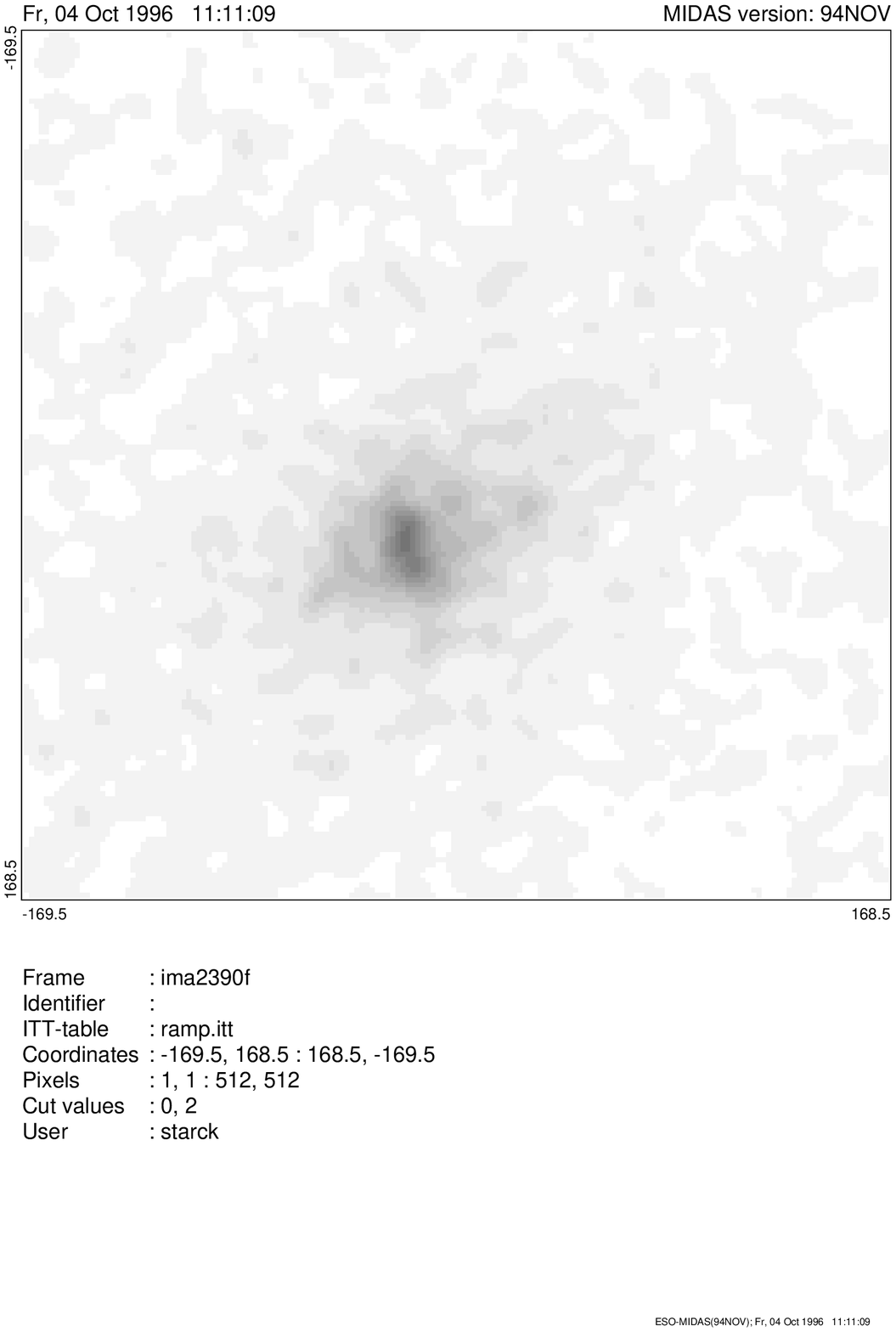,bbllx=3.1cm,bblly=10.5cm,bburx=18.7cm,bbury=26.2cm,width=10cm,height=10cm,clip=}
}}
\caption{A2390 ROSAT image filtered by a standard method (convolution with a Gaussian).}
\label{fig_ima_fg_a2390_col}
\end{figure*}

\begin{figure*}[htb]
\centerline{
\vbox{
\psfig{figure=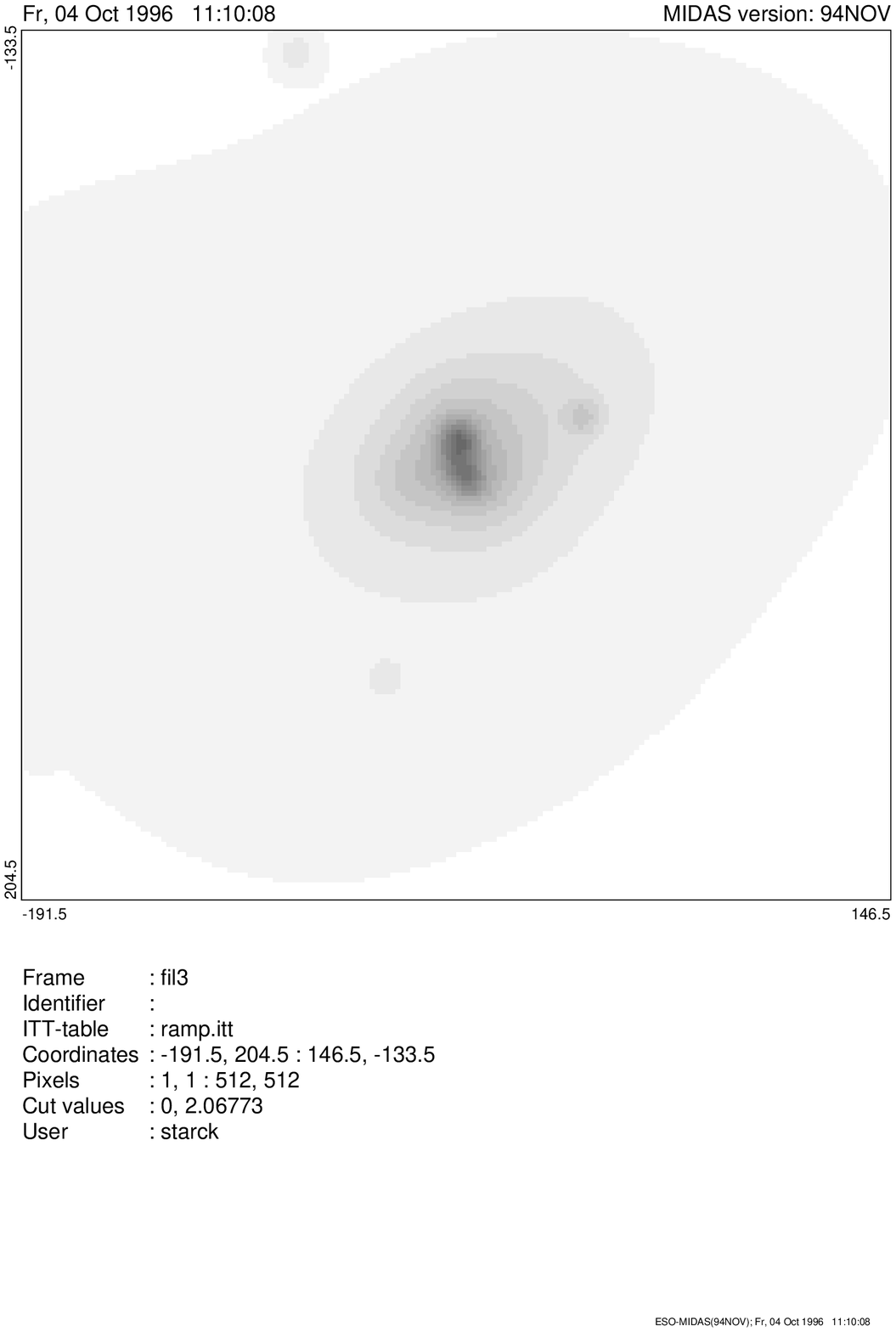,bbllx=3.1cm,bblly=10.5cm,bburx=18.7cm,bbury=26.2cm,width=10cm,height=10cm,clip=}
}}
\caption{A2390 ROSAT image filtered by the wavelet based method.}
\label{fig_ima_fw_a2390_col}
\end{figure*}

The cluster of galaxies A2390 is located at a redshift of 0.231.
Figure~\ref{fig_ima_a2390_bw} shows an image of this cluster, obtained
with   ROSAT satellite. The resolution is one arc second per pixel,
with a total number of 13506 photons for exposure time of approximately
8 hours. The background level is around  0.04 photons per pixel.
It is clear that the raw data are not usable, and we need to filter it 
in order to extract the information. The standard method consists in 
convolving the image by a Gaussian. Figure~\ref{fig_ima_fg_a2390_col} 
shows the result after applying this convolution (Gaussian with a
full width at half maximum equal to 5", which is approximatively the
size of the instrumental response). The smoothed image shows structure,
but we see also that a lot of noise remains, and it is difficult to 
assign a significance to these structures. Figure~\ref{fig_ima_fw_a2390_col}
shows the filtered image by the histogram based wavelet method. The noise
has been eliminated, and we see that the wavelet transform has  
 enhanced weak structures in the X-ray emission, which could explain 
the gravitational amplification phenomena which have been observed in
the optical domain (Pierre et al. 1996).

\section{Conclusion}
Simulations have shown that the best filtering approach for images containing
Poisson noise with few events is the method based on the histogram autoconvolutions.
This method allows one to give a probability that a wavelet coefficient is due
to noise. No background model is needed, and simulations with different 
background levels have shown the reliability and the robustness of the
method. Other noise models in the wavelet space lead to the problem of 
the  significance of the wavelet coefficient. A ten sigma detection
was not strong enough in our simulation to produce a good filtered image. 
In this case,  only Monte Carlo simulations can allow one  derivation of a good 
detection level, and then, a new problem appears of defining the
correct background. The main advantage of the histograms based method
is its independence of the background. 

\acknowledgements
We wish to thank A.~Bijaoui and E.~Slezak for useful discussions and comments,
and C.~Delattre for his technical help.

\end{document}